\documentclass[twocolumn,amsmath,amssymb,superscriptaddress,aps,pra]{revtex4}
\usepackage{amsmath,amsthm,amssymb}
\usepackage{latexsym}
\usepackage{amscd,graphicx}
\usepackage[titletoc]{appendix}
\usepackage{epsfig}
\usepackage{epstopdf}
\usepackage[normalem]{ulem}
\usepackage[dvipsnames]{xcolor}
\usepackage[colorlinks=true,linkcolor=red,citecolor=blue]{hyperref}
\usepackage{csquotes}
\usepackage{graphicx}
\usepackage{adjustbox}
\usepackage{tabularx} 

\usepackage{booktabs}

\begin{document}

\title{Hybrid Cavity-Magnon Optomechanics: Tailoring Bipartite and Tripartite Macroscopic Entanglement}

\author{Qin-Geng Chen}
\affiliation{Department of Physics, Wenzhou University, Zhejiang 325035, China}

\author{Ming-Yue Liu}
\affiliation{Department of Physics, Wenzhou University, Zhejiang 325035, China}

\author{Xian-Xian Huang}
\affiliation{Department of Physics, Wenzhou University, Zhejiang 325035, China}

\author{Jiaojiao Chen}
\altaffiliation{jjchenphys@hotmail.com}
\affiliation{Department of Physics, Wenzhou University, Zhejiang 325035, China}

\author{Wei Xiong}
\altaffiliation{xiongweiphys@wzu.edu.cn}
\affiliation{Department of Physics, Wenzhou University, Zhejiang 325035, China}
\affiliation{International Quantum Academy, Shenzhen, 518048, China}

\date{\today }

\begin{abstract}
Cavity optomechanics, providing an inherently nonlinear interaction between photons and phonons, have shown enomerous potential in generating macroscopic quantum entanglement. Here we propose to realize diverse bipartite and tripartite entanglement in cavity-magnon optomechanics. By introducing magnons to standard cavity optomechanics, not only tunable optomechanical entanglement and magnon-magnon entanglement can be achieved, but also flexible tripartite entanglement including magnon-photon-phonon entanglement, magnon-magnon-photon and -phonon entanglement can be generated. Moreover, optimal bipartite and tripartite entanglement can be achieved by tuning parameters. We further show that all entanglement can be enhanced via engineering the magnon-photon coupling, and is proven to be robust against the bath temperature within the survival temperature. Besides, we find that the optomechanical entanglement can be protected or restored by bad magnons with large decay rate, while other entanglement is severely reduced. The results indicate that our proposal provides a novel avenue to explore and control tunable macroscopic quantum effects in hybrid cavity-magnon optomechanics.
\end{abstract}


\maketitle

\section{introduction}

Magnons~\cite{prabhakar2009spin,Kittel1958}, quanta of collective spin-wave excitations in magnetically ordered materials like yttrium iron garnet (YIG), have garnered considerable interest in quantum physics and condensed matter physics~\cite{rameshti2022cavity,yuan2022quantum,zheng2023tutorial,li2020hybrid,lachance2019hybrid,zuo2024cavity}. With unique properties such as high spin density and low energy loss
, diverse novel phenomena including strong~\cite{huebl2013high,tabuchi2014hybridizing,zhang2014strongly,PhysRevAppliedHigh} and ultrastrong coupling~\cite{ghirri2023ultrastrong,wang2024ultrastrong,kostylev2016superstrong}, qubit-magnon coupling~\cite{tabuchi2015coherent}, magnon dark modes and gradient memory~\cite{zhang2015magnon}, magnon Kerr effect~\cite{wang2016magnon}, spin current manipulation~\cite{bai2017cavity}, exceptional points~\cite{zhang2017observation,zhang2019higher,PhysRevB.95.214411,liu2019observation}, axion detection~\cite{PhysRevD.102.095005,PhysRevD.105.102004,BARBIERI1989357}, superadiant phase transition~\cite{liu2023switchable}, dissipative coupling~\cite{wang2020dissipative,harder2021coherent}, and strong spin coupling~\cite{xiong2022strong,xiong2023optomechanical,peng2025cavity,PhysRevLett.125.247702,PhysRevLett.130.073602,PRXQuantum.2.040314,PhysRevB.106.235409,chen2025exponentially}. Very recently, a single magnon manipulation~\cite{xu2023quantum,xu2024macroscopic} and quantum magnon squeezing~\cite{you2025quantum} have been demonstrated. This greatly promotes the further development of magnonics. In magnonics, how to prepare macroscopic entanglement is a long outstanding problem~\cite{li2018magnon,zhang2019quantum,PhysRevB.108.024105,PhysRevA.109.043512,yuan2020enhancement,li2019entangling,sun2021remote,ren2022long,lai2022noise,liu2025nonreciprocal,PhysRevLett.124.213604,PhysRevApplied.15.024042,PhysRevResearch.3.013192,PhysRevB.106.224404,noura2024enhanced}. This is because it is crucial for understanding the classical-quantum boundary~\cite{haroche1998entanglement} and can be regarded as an important resource for quantum information science~\cite{bouwmeesterphysics}. However, to produce such macroscopic entanglement, nonlinear effects are always prerequired~\cite{adesso2007entanglement}. The extensively explored cavity optomechanics~\cite{RevModPhys.86.1391,kippenberg2007cavity,kippenberg2008cavity,metcalfe2014applications,liu2013review,chang2022cavity,barzanjeh2022optomechanics,xiong2015review,amazioug2020enhancement}, hybridized by photons in a cavity mode weakly coupled to phonons in a mechanical resonator through radiation pressure~\cite{vitali2007optomechanical}, can provide the natural nonlinearity for investigating optomechanical entanglement~\cite{sarma2021continuous} and other intriguing effects such as sensing~\cite{li2021cavity,liu2021progress,schliesser2008high}, ground-state cooling~\cite{marquardt2008quantum,otterstrom2018optomechanical,gan2019intracavity,park2009resolved,xu2019nonreciprocal,liu2014optimal,huang2022multimode,genes2008ground,PhysRevA.101.063836,nunnenkamp2010cooling,fan2025hybrid}, squeezed light generation~\cite{purdy2013strong,aggarwal2020room,kronwald2014dissipative,lu2015steady,liao2011parametric,lu2015squeezed}, nonreciprocity~\cite{PhysRevLett.125.023603,PhysRevLett.130.013601,Xu:20,PhysRevA.106.032606,Hafezi:12,PhysRevX.7.031001,PhysRevApplied.7.064014,PhysRevLett.125.143605,PhysRevA.98.063845}, optomechanically induced transparency~\cite{PhysRevA.101.043820,PhysRevApplied.10.014006,PhysRevLett.111.133601,PhysRevA.87.055802,PhysRevA.98.053802,PhysRevA.102.023707,PhysRevLett.130.093603,PhysRevA.88.013804,PhysRevLett.110.223603,weis2010optomechanically}, coupling enhancement~\cite{PhysRevB.103.174106,lu_quantum-criticality-induced_2013,Chen:21,shen2016experimental,fang2017generalized}, multistability~\cite{PhysRevA.93.023844,PhysRevLett.112.076402,PhysRevA.85.043824,PhysRevA.84.033846}, and high-order exceptional points~\cite{xiong_higher-order_2022,xiong_higher-order_2021,jing2017high}. This indicates that combination of magnonics and optomechanics is a potential path to investigate macroscopic magnon entanglement.

Owing to the strong photon-magnon coupling in planar~\cite{huebl2013high} and 3D cavities~\cite{tabuchi2014hybridizing,zhang2014strongly,PhysRevAppliedHigh}, hybridizing magnons with optomechanics to form cavity-magnon optomechanics becomes feasible~\cite{PhysRevLett.129.243601,PhysRevA.109.043512,PhysRevA.96.023826,PhysRevB.108.024105,xiong2023optomechanical,sohail2025coherent,mathkoor2025bipartite,hidki2024entanglement,ahmed2025nonreciprocal} for utilizing their individual advantages. In this work, we propose to generate diverse bipartite and tripartite entanglement in hybrid cavity-magnon optomechanics. Compared to the standard cavity optomechanics, the optomechanical entanglement can be adjusted flexibly by the introduced magnons in the Kittle mode of a single sphere. Specifically, the optimal optomechanical entanglement can be predicted by tuning the detuning of the Kittle mode from the driving field to be resonant with the mechanical mode (i.e., $\Delta_m=\omega_b$). But when we tune this detuning to be anti-resonant with the mechanical mode (i.e., $\Delta_m=-\omega_b$), the optomechanical entanglement is weakest, while the magnon-photon entanglement, the magnon-phonon, as well as the magnon-photon-phonon entanglement are approximately optimal. These entanglement can be further enhanced by engineering the coupling between magnons and photons. In addition, the tunable magnon-magnon entanglement and the tripartite entanglement including the magnon-magnon-photon entanglement and the magnon-magnon-phonon entanglement can also be generated in the proposed cavity-magnon optomechanics with two spheres. Tuning one Kittle mode to be resonant with the mechanical mode (i.e., $\Delta_1=\omega_b$), while the other to be anti-resonant with the mechanical mode (i.e., $\Delta_2=-\omega_b$), or vice versa, the optimal magnon-magnon entanglement is predicted, and its value approximately equals to half the optomechanical entanglement. But to have optimal magnon-magnon-photon and -phonon entanglement, two Kittle modes are needed to be resonant (i.e., $\Delta_1=\Delta_2$). We further show that the optomechanical entanglement can be protected or restored by using bad magnons with large decay rate, but the magnon-magnon entanglement as well as the magnon-magnon-photon and -phonon entanglement are severly reduced. Besides, the bath temperature effect on these entanglement are also investigated. We find that for the weaker magnon-photon coupling (i.e., $g_m<G$), the stronger optomechanical entanglement than that of the magnon-magnon entanglement can be obtained within the survival temperature, but the situation is reversed when the magnon-photon coupling becomes comparable with the optomechanical coupling (i.e., $g_m\approx G$). Moreover, the survival temperature of the magnon-magnon entanglement is more robust against the magnon-photon coupling than that of the optomechanical entanglement. For the tripartite entanglement, we show that the magnon-magnon-phonon entanglement is stronger than the magnon-{magnon}-photon entanglement within their survival temperature, owing to the mechanical cooling. All achieved entanglement exhibit {robustness} against the bath temperature. Our proposal opens an alternative path to study tunable macroscopic quantum effect in hybrid cavity-magnon optomechanics.

{The rest of the paper} is organized as follows: In Sec.~\ref{s2}, the model
is described and the entanglement quantification is given.
Then in Sec.~\ref{s3}, we study optomechanical entanglement mediated by magnons in the proposed cavity-magnon optomechanics, where only a single sphere is considered. In Sec.~\ref{s4}, tunable magnon-magnon entanglement induced by the optomechanical interface is investigated via examing the effect of system parameters. In Sec.~\ref{s5}, we further study tripartite entanglement including magnon-magnon-photon and magnon-magnon-phonon entanglement.  Finally, a conclusion is given in Sec.~\ref{s6}.

\section{Model and Entanglement Quantification}\label{s2}
\subsection{Model}
\begin{figure}
	\includegraphics[scale=0.5]{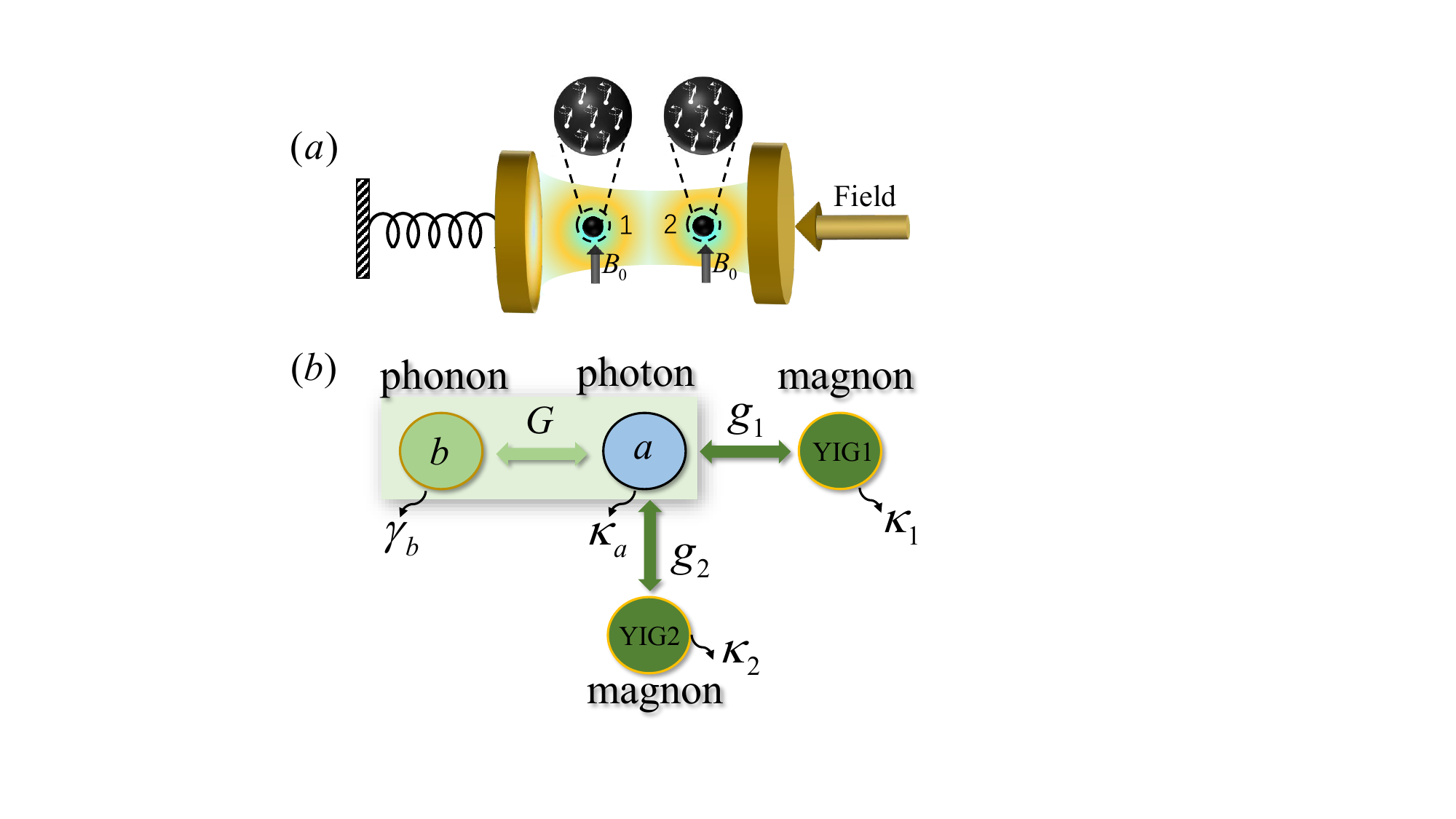}
	\caption{(a) Schematic diagram of the proposed cavity-magnon optomechanics. It consists of two YIG spheres placed in biased magnetic field $B_0$, a driven cavity and a mechanical resonator. The cavity mode is strongly coupled to the Kittle modes of two YIG spheres, and weakly coupled to the mechanical mode. (b) The effective coupling configuration. $G$ is the linearized optomechanical coupling strength, and $g_{1(2)}$ is the tunable coupling strength between the cavity mode and the Kittle mode of the YIG sphere $1~(2)$.}\label{fig1}
\end{figure}

We consider a hybrid cavity-magnon optomechanics consisting of a mechanical resonator weakly coupled to a cavity containing two separated YIG spheres, as depicted in Fig.~\ref{fig1}(a). The magnons in the Kittle modes of two spheres are strongly coupled to the photons in the cavity mode with the uniform coupling strength $g_1=g_2=g_m$. By imposing an external field, with the frequency $\omega_0$ and the amplitude $\Omega$, to the cavity, the Hamiltonian of the proposed hybrid system with respect to $\omega_0$, can be given by (setting $\hbar=1$)
\begin{align}\label{eq1}
	H_{\rm sys}=&H_{\rm OM}+\sum\limits_{j=1,2}H_{\rm CM}^{(j)}+i \Omega\left(a^{\dagger}-a\right),
\end{align}
where $H_{\rm OM}=\Delta_a a^{\dagger} a+\frac{\omega_b}{2}\left(q^2+p^2\right)+g_0 a^{\dagger} a q$ denotes the {Hamiltonian} of the standard cavity optomechanics, describing the interaction between the cavity mode and the mechanical mode via the radiation pressure, with the single-photon optomechanical coupling strength $g_0$. $\Delta_a=\omega_a-\omega_0$, with $\omega_a$ being the frequency of the cavity mode, is the frequency detuning of the cavity mode from the driving field, and $\omega_b$ is the frequency of the mechanical mode. The Hamiltonian $H_{\rm CM}^{(j)}=\Delta_j m^{\dagger}_j m_j+g_j(m_j a^\dag+m_j^\dag a)$ charaterizes the interaction between the cavity mode and the Kittle mode of the $j$th sphere. $\Delta_j=\omega_j-\omega_0$ is the frequency detuning of the Kittle mode of the $j$th sphere from the driving field. The creation (annihilation) operators the cavity mode and the Kittle mode of the $j$th sphere are denoted by $a^\dagger$ ($a$) and $m_j^\dagger$ ($m_j$), respectively. $q$ and $p$ are dimensionless position and momentum quadratures of the mechanical mode.

\subsection{Quantum Langevin equation}

By taking input noises and dissipations into account, the dynamics of the proposed system can be governed by the quantum Langevin equations, i.e.,
\begin{align}\label{eq2}
		 \dot{q}=&\omega_b p,\notag\\
		 \dot{p}=&-\omega_b q-g_0 a^{\dagger} a-\gamma_b p+\xi(t),\notag\\ 
		  \dot{m_j}=&-(\kappa_j+i \Delta_j) m_j-i g_j a+\sqrt{2 \kappa_j} \zeta_j(t),\\
		  \dot{a}=&-(\kappa_a+i{\Delta}_a) a-ig_0 aq-ig_j m_j+\Omega+\sqrt{2 \kappa_a} \mu(t),\notag
\end{align}
{where $\kappa_a$, $\gamma_b$, and $\kappa_{m,j}$ denote the decay rates of the cavity mode, the mechanical mode, and the Kittel mode of the $j$th YIG sphere, respectively. 
The corresponding vacuum input noise operators are $\mu(t)$, $\xi(t)$, and $\zeta_j(t)$, which satisfy zero mean values 
$\langle \mu(t)\rangle = \langle \xi(t)\rangle = \langle \zeta_j(t)\rangle = 0$. 
Under the Markovian approximation, their two-time correlation functions are given by
\begin{align}\label{eq3}
	\langle \mu(t) \mu^\dagger(t^{\prime})\rangle &= (\bar{n}_a+1)\delta(t-t^{\prime}), \notag\\
	\langle \mu^\dagger(t) \mu(t^{\prime})\rangle &= \bar{n}_a \delta(t-t^{\prime}), \notag\\
	\langle \zeta_j(t) \zeta_j^\dagger(t^{\prime})\rangle &= (\bar{n}_j+1)\delta(t-t^{\prime}), \\
	\langle \zeta_j^\dagger(t) \zeta_j(t^{\prime})\rangle &= \bar{n}_j \delta(t-t^{\prime}), \notag\\
	\langle \xi(t)\xi(t^{\prime})+\xi(t^{\prime})\xi(t)\rangle &\simeq 2\gamma_b(2\bar{n}_b+1)\delta(t-t^{\prime}), \notag
\end{align}
where $\bar{n}_\sigma = [\exp(\hbar \omega_\sigma / k_B T) - 1]^{-1}$ ($\sigma = a, b, j$) is the mean thermal excitation number, $k_B$ is the Boltzmann constant, and $T$ is the bath temperature.}

In fact, Eq.~(\ref{eq2}) can be linearized when the external driving field is strong. For this, we first rewrite the operator $O=\langle O\rangle+\delta O$ ($O=q,p,a,m_j$), where $\langle O\rangle$ is the steady-state value and $\delta O$ is the fluctuation operator. Then we substitue the redefined operator $O$ into Eq.~(\ref{eq2}) and neglect the high-order fluctuation terms,  resulting in
	\begin{align}\label{eq4}
		\dot{\delta q}=&\omega_b \delta p,\notag\\
		\dot{\delta p}=&-\omega_b \delta q-(G^*\delta a+G \delta a^\dag)/\sqrt{2}-\gamma_b \delta p+\xi,\notag\\ 
		\dot{\delta m_j}=&-(\kappa_j+i \Delta_j) \delta m_j-i g_j \delta a+\sqrt{2 \kappa_j} \zeta_j,\\
		\dot{\delta a}=&-(\kappa_a+i\tilde{\Delta}_a) \delta a-iG q/\sqrt{2}-ig_j\delta m_j+\sqrt{2 \kappa_a} \mu,\notag
	\end{align}
where $\tilde{\Delta}_a={\Delta}_a+g_0\langle q\rangle$ is the effective frequency detuning, induced by the displacement of the mechanical mode, and $G=\sqrt{2}g_0\langle a\rangle$ is the effectively enhanced optomechanical coupling strength between the cavity mode and the mechanical mode by the factor $\langle a\rangle$. The steady-state values $\langle O\rangle$ in Eq.~(\ref{eq4}) are given by $\dot{\langle O\rangle}=0$. Specifically, $\langle p\rangle=0$, $\langle q\rangle=-{g_0|\langle a\rangle|^2}/{\omega_b}$, $\langle m_j\rangle\approx- {g_j\langle a\rangle}/{\Delta_j}$, and $	\langle a\rangle \approx {i \Omega}/{(g_j^2/ \Delta_j-\tilde{\Delta}_a)}$, where $\left|\Delta_j\right|,\left|\tilde{\Delta}_a\right|\gg\kappa_j,\kappa_a$ are taken. Experimentally, these conditions can be easily achieved by tuning the frequency of the driving field~\cite{wang2016magnon}. Also, the value of $\langle a\rangle$ can be real via tuning the initial phase of the strong driving field. Hereafter, we assume $\langle a\rangle$ to be real for simplicity{, directly giving rise to $G=G^*$}. Thus, the effective Hamiltonian of the linearized hybrid system can be written as
\begin{align}\label{eq6}
H_{\rm eff}=&\tilde{\Delta}_a a^\dag a+\frac{\omega_b}{2}\left(q^2+p^2\right)+\frac{G}{\sqrt{2}}(a+a^\dag)q\notag\\
&+\sum\limits_{j=1,2}\Delta_j m_j^\dag m_j+g_j(m_j a^\dag+a m_j^\dag),
\end{align}
which effectively describes the interaction among the cavity mode, the mechanical mode, and the Kittle modes, as illustrated in Fig.~\ref{fig1}(b). Notably, all parameters in the Hamiltonian~(\ref{eq6}) are adjustable except for the mechanical frequency $\omega_b$. The fluctuation symbol \enquote{$\delta$} is omitted for convenience.
\subsection{Covariance matrix}
By further defining fluctuation quadratures of the Kittle and the cavity modes as
\begin{align}
	x_j=&(m_j+m_j^{\dagger}) / \sqrt{2},~~y_j=i(m_j^{\dagger}-m_j) / \sqrt{2},\notag\\
	 x_a=&(a+a^{\dagger}) / \sqrt{2},~~y_a=i(a^{\dagger}-a) / \sqrt{2},
\end{align}
and the associated input noise operators as
\begin{align}
	 X_{j}=&(\zeta_j+\zeta_j^{\dagger}) / \sqrt{2},~~Y_j=i(\zeta_j^{\dagger}-\zeta_j) /\sqrt{2},\notag\\
	 X_\mu=&(\mu+\mu^{\dagger}) / \sqrt{2},~~Y_\mu=i(\mu^{\dagger}-\mu) / \sqrt{2},
\end{align}
the dynamics in Eq.~(\ref{eq4}) can be equivalently written as the matrix form,
\begin{align}
	\dot{u}(t)=A u(t)+n(t),
\end{align}
where $u(t)=[x_1,y_1, x_2,y_2, x_a,y_a,q,p]^T$ is the vector operator of the system, $n(t)=[\sqrt{2\kappa_1}X_1,\sqrt{2\kappa_1}Y_1,\sqrt{2\kappa_2}X_2,\\\sqrt{2\kappa_2}Y_2,\sqrt{2\kappa_a}X_\mu,\sqrt{2\kappa_a}Y_\mu,0,\xi]^T$ is the vector operator of the input noise, and

\begin{align}
		A=\left(\begin{array}{cccccccc}
			-\kappa_1 & \Delta_1 & 0 & 0 & 0 & g_1 & 0 & 0 \\
			-\Delta_1 & -\kappa_1 & 0 & 0 & -g_1 & 0 & 0 & 0 \\
			0 & 0 & -\kappa_2 & \Delta_2 & 0 & g_2 & 0 & 0 \\
			0 & 0 & -\Delta_2 & -\kappa_2 & -g_2 & 0 & 0 & 0 \\
			0 & g_1 & 0 & g_2 & -\kappa_a & \tilde{\Delta}_a & 0 & 0 \\
			-g_1 & 0 & -g_2 & 0 & -\tilde{\Delta}_a & -\kappa_a & -G & 0 \\
			0 & 0 & 0 & 0 & 0 & 0 & 0 & \omega_b \\
			0 & 0 & 0 & 0 & -G & 0 & -\omega_b & \gamma_b
		\end{array}\right)
\end{align}
is the drift matrix. According to the Routh-Hurwitz criteria\cite{Hurwitz1964,PhysRevA.35.5288}, the system is stable only when all the eigenvalues of $A$ have negative real parts. The stability is ensured numerically in the following.

Since the input quantum noises are zero-mean quantum Gaussian noises, the quantum steady state for the
fluctuations is a zero-mean continuous variable Gaussian state, fully characterized by an $8\times8$ covariance matrix $\mathcal{V}_{lk}(t)=\frac{1}{2}\left\langle u_l(t) u_k\left(t^{\prime}\right)+u_k\left(t^{\prime}\right) u_l(t)\right\rangle(l,k=1,2, \ldots, 8)$. The matrix $\mathcal{V}$ can be obtained 
by directly solving the Lyapunov equation\cite{Stability1993} 
\begin{align}
A \mathcal{V}+\mathcal{V} A^T+\mathcal{D} =0,\label{eq8}
\end{align}
where $\mathcal{D}=\operatorname{diag}[\kappa_1(2 \bar{n}_1+1), \kappa_1(2 \bar{n}_1+1),\kappa_2(2 \bar{n}_2+1), \kappa_2(2 \bar{n}_2+1),\kappa_a(2 \bar{n}_a+1),\kappa_a(2 \bar{n}_a+1), 0, \gamma_b(2 \bar{n}_b+1)]$
is defined by $\langle \bar{n}_l(t) \bar{n}_k(t^{\prime})$ $+\bar{n}_k(t^{\prime}) \bar{n}_l(t)\rangle=2\mathcal{D}_{lk} \delta(t-t^{\prime})$.

\subsection{Bipartite and tripartite entanglement}

Once the matrix $\mathcal{V}$ is obtained by solving Eq.~(\ref{eq8}), arbitrary bipartite entanglement can be quantified by the logarithmic negativity (LN)~\cite{PhysRevLett.95.090503},
\begin{align}\label{eq10}
	E_N\equiv \max [0,-\ln 2{\eta^-}]
\end{align}
with
\begin{align}\label{eq21}
	\eta^-=2^{-1/2}[\Sigma-(\Sigma^2-4\rm det {V}_4)^{1/2}]^{1/2},
\end{align}
where $ \Sigma=\rm{det} \mathcal{A}+\rm{det} \mathcal B-2\rm{det} \mathcal C $ and 
$\mathcal{V}_4=\begin{pmatrix}
	\mathcal A &\mathcal C\\
	\mathcal {C^T}& \mathcal B\\
\end{pmatrix}$
is the $4 \times4 $ block form of the correlation matrix, associated with two modes of interest~\cite{PhysRevLett.84.2726}. $\mathcal{A}$, $\mathcal{B}$, and $\mathcal{C}$ are the $ 2\times2 $ blocks of $\mathcal{V}_4$. $E_N>0$ means that the interested two modes are entangled. 

Additionally, tripartite entanglement among arbitrary three interested modes can also be studied by the function of the minimal residual contangle (MRC)~\cite{Adesso_2007,Adesso_2006}:
\begin{align}
	\mathcal{R}_\tau \equiv \min \left[\mathcal{R}_\tau^{r \mid s t}, \mathcal{R}_\tau^{s \mid r t}, \mathcal{R}_\tau^{t \mid r s}\right],
\end{align}
where
\begin{align}
	\mathcal{R}_\tau^{r \mid s t} \equiv C_{r \mid s t}-C_{r \mid s}-C_{r \mid t},
\end{align}
with $C_{u \mid v}$ being the contangle of a subsystem of $u$ and $v$ ($v$ contains one
or two modes) and $\{r,s,t\}\in\{a,b,m_1,m_2\}$, is a proper entanglement monotone defined as the squared LN~\cite{Adesso_2007,Adesso_2006}. $\mathcal{R}_\tau> 0$ indicates that the interested three modes are entangled.

\section{Optomechanical entanglement}\label{s3}

In this part, we study the magnon-mediated optomechanical entanglement. To be more clear, we first re-examine the optomechanical entanglement without magnons, i.e., $g_1=0$ and $g_2=0$. Then we investigate the magnon-mediated optomechanical entanglement by taking a single sphere for example. Without loss of {generality}, we assume $g_1=g_m\neq0$ and $g_2=0$.

\begin{table}[htbp]
	\centering
	\caption{\textcolor{red}{Symbols used in numerical calculation}}
	\label{table2}
	\begin{adjustbox}{width=0.8\columnwidth}
		\begin{tabular}{ll}
			\hline\hline
			\textbf{Symbol} & \textbf{Meaning} \\
			\hline
			$\omega_a/2\pi = 10~\mathrm{GHz}$ & Cavity frequency \\
			$\omega_b/2\pi = 10~\mathrm{MHz}$ & Mechanical frequency \\
			$\omega_{1(2)}/2\pi = 10~\mathrm{GHz}$ & Magnon frequency \\
			$\kappa_a/2\pi = 1~\mathrm{MHz}$ & Cavity decay rate \\
			$\gamma_b/2\pi = 100~\mathrm{Hz}$ & Mechanical decay rate \\
			$\kappa_{1(2)}/2\pi = 1~\mathrm{MHz}$ & Magnon decay rate \\
			$\omega_0/2\pi = 9.95~\mathrm{GHz}$ & Frequency of the driving field \\
			$\Omega/2\pi = 0.4~\mathrm{GHz}$ & Amplitude of the driving field \\
			$T = 20~\mathrm{mK}$ & Bath temperature \\
			$\tilde{\Delta}_m = 0.9\omega_b$ & Cavity detuning \\
			$\Delta_{1(2)} = 5\omega_b$ & Magnon detuning \\
			\hline\hline
		\end{tabular}
	\end{adjustbox}
\end{table}

\subsection{Without magnons}

\begin{figure}
	\includegraphics[scale=0.265]{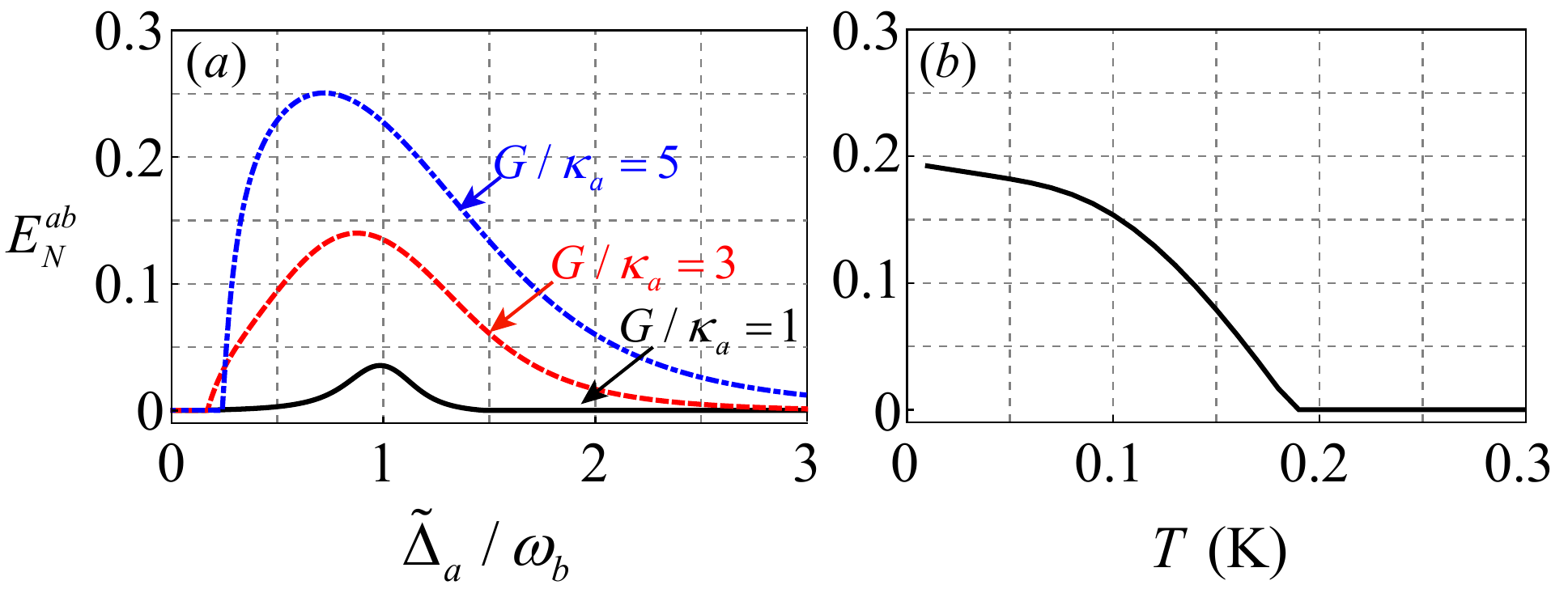}  
	\caption{(a) The optomechanical entanglement ($E_N^{ab}$) vs the {dimensionless} normalized cavity detuning $\tilde{\Delta}_a/\omega_b$ with different optomechanical coupling strengths $G/\kappa_a=1,3,5$ at the bath temperature $T=20$ mK{, where $\tilde{\Delta}_a$ is the cavity detuning and $\omega_b$ is the frequency of the mechanical mode}. (b) The LN $E_N^{ab}$ vs the bath temperature $T$ with the moderate optomechanical coupling strength $G/\kappa_a=4$ and the frequency detuning $\tilde{\Delta}_a=0.9\omega_b$. Other parameters are chosen as $\omega_a/2\pi=10$ GHz, $\omega_b/2\pi=10$ MHz, $\kappa_a/2\pi=1$ MHz, $\gamma_b/2\pi=100$ Hz.}\label{fig2}
\end{figure}
When no spheres are placed in the cavity (i.e., $g_1=g_2=0$), the proposed cavity-magnon optomechanics reduces to a standard optomechanical configuration comprising solely a cavity and a mechanical resonator. In this setup, optomechanical entanglement can be analyzed by plotting LN ($E_N^{ab}$) vs the normalized cavity detuning $\tilde{\Delta}_a/\omega_b$ with different linearized coupling strengths $G$ in Fig.~\ref{fig2}(a). Here, we employ experimentally accessible parameters\cite{zhang2016cavity, bernier2017nonreciprocal}: $\omega_a/2\pi$ = 10 GHz, $\omega_b/2\pi$ = 10 MHz, $\kappa_a/2\pi=1$ MHz, $\gamma_b/2\pi=10^2$ Hz, and $T=20$ mK. { Besides, other parameters such as $\omega_0=9.95$ GHz and $\Omega=0.4$ GHz are chosen.} Evidently, the optomechanical entanglement increases first and then decreases with the cavity detuning $\tilde{\Delta}_a/\omega_b$. Around $\tilde{\Delta}_a\approx\omega_b$, the optimal optomechanical entanglement is observed. By gradually increasing the optomechanical coupling, the optomechanical entanglement can be significantly enhanced (as depicted by the three curves). The generated optomechanical entanglement arises from both the two-mode squeezing effect ($a^\dag b^\dag+ab$, where $q\propto b+b^\dag$) and the beam-splitter effect ($a^\dag b+ab^\dag$). While the former generates entanglement, and the latter cools the mechanical resonator to sustain the entanglement. In Fig.~\ref{fig2}(b), the effect of the bath temperature on the optomechanical entanglement is further elucidated by fixing $\tilde{\Delta}_a=0.9\omega_b$ and $G=4\kappa_a$. It is shown that the optomechanical entanglement monotonically decreases as the bath temperature increases. The survival temperature for the optomechanical entanglement can be up to $\sim 190$ mK.

\begin{figure}
	\includegraphics[scale=0.3]{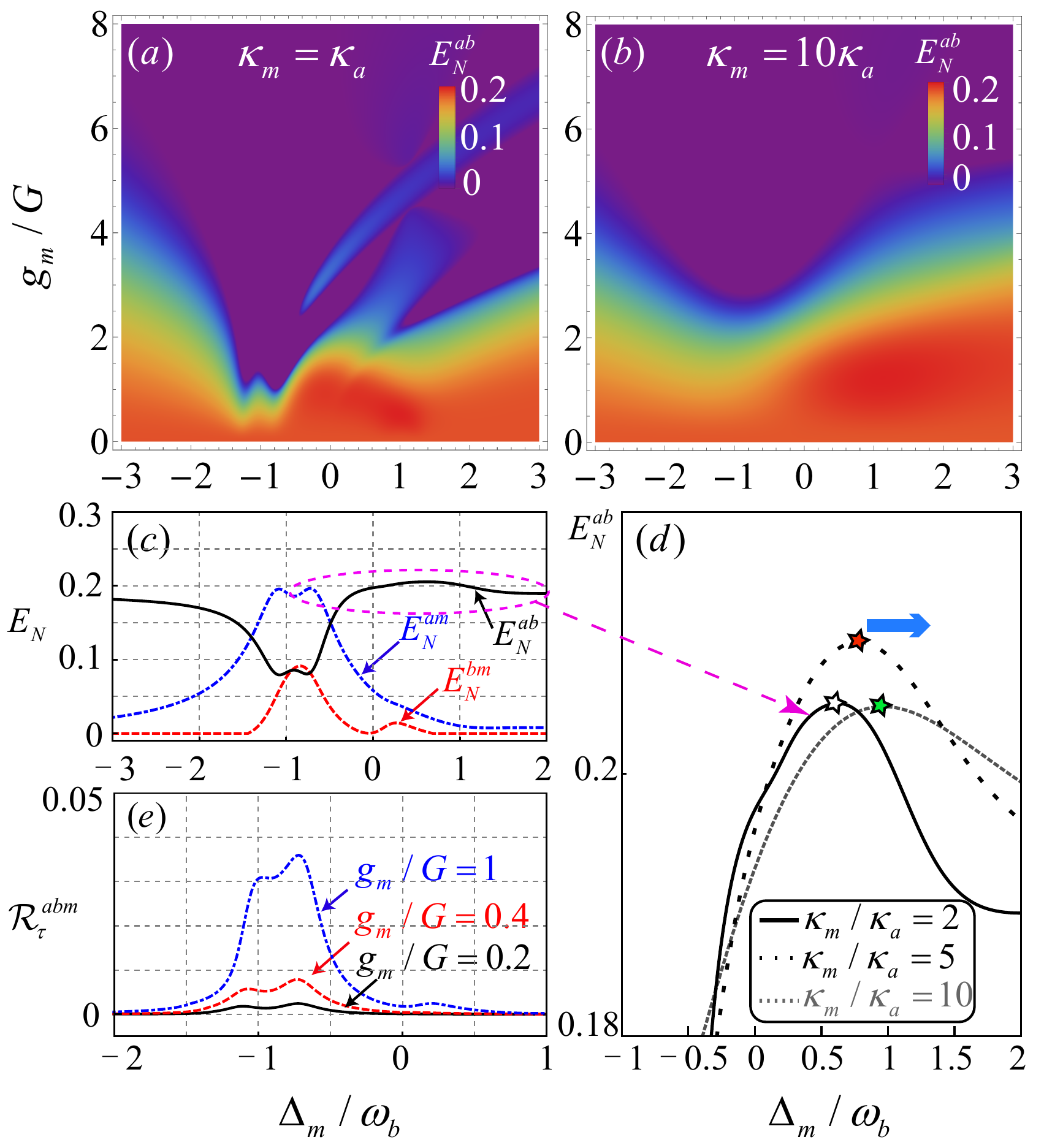}
	\caption{Density plot of $E_N^{ab}$ vs the normalized frequency detuning $\Delta_m/\omega_b$ and the normalized magnon-photon coupling strength $g_m/G$ with (a) $\kappa_m=\kappa_a$ and (b) $\kappa_m=10\kappa_a$. (c) The optomechanical entanglement ($E_N^{ab}$), the magnon-photon entanglement ($E_N^{am}$), and the magnon-phonon entanglement ($E_N^{bm}$), vs the normalized magnon frequency detuning $\Delta_m/\omega_b$, where $g_m=G$ and $\kappa_m=2\kappa_a$. {(d) The optomechanical entanglement vs the normalized frequency detuning $\Delta_m/\omega_b$ with different values of $\kappa_m/\kappa_a=2,5,10$. The star means the optimal entanglement. (e) The magnon-photon-phonon entanglement ($\mathcal{R}_\tau^{abm}$) vs the normalized magnon frequency detuning $\Delta_m/\omega_b$ with different magnon-photon coupling strengths $g_m/G=0.2,0.6,1$, where $\kappa_m=\kappa_a$ is taken.} Other parameters are the same as those in Fig.~\ref{fig2}(a) except for $G=4\kappa_a$ and $\tilde{\Delta}_a=0.9\omega_b$.}\label{fig3}
\end{figure}

\subsection{With magnons}

{To investigate the magnon mediated optomechanical entanglement, we take the hybrid cavity-magnon optomechanical system with a single YIG sphere as an example, that is, we set $g_1 = g_m \neq 0$ and $g_2 = 0$ in Fig.~\ref{fig1}, and accordingly specify $\Delta_1 = \Delta_m$. Within the framework of cavity magnonics, parameters such as the magnon frequency, the magnon-photon coupling strength, and the magnon decay rate can be tuned, resulting in controllable optomechanical entanglement. The corresponding numerical results are shown in Fig.~\ref{fig3}. Figure~\ref{fig3}(a) shows how the optomechanical entanglement varies with the magnon-photon coupling strength $g_m$ while keeping other system parameters fixed. The entanglement gradually decreases as $g_m$ increases. Nevertheless, an optimal entanglement value appears when $g_m$ becomes comparable to the effective optomechanical coupling $G$, that is, $g_m \sim G$. This optimal behavior is further confirmed in Fig.~\ref{fig3}(b), which demonstrates that the optomechanical entanglement reaches its maximum around the same parameter range. Moreover, a comparison between Figs.~\ref{fig3}(a) and \ref{fig3}(b) reveals that a larger magnon decay rate can help preserve or even recover the optomechanical entanglement, highlighting the nontrivial role of the dissipative magnons in the hybrid system. The effect of the magnon decay rate is further examined in Fig.~\ref{fig3}(c). It is shown that introducing magnons with a small decay rate leads to a noticeable reduction in the optomechanical entanglement. This reduction is accompanied by a redistribution of quantum correlations into magnon–photon (blue curve) and magnon-phonon (red curve) entanglements, particularly when the cavity detuning approaches $\Delta_m \approx -\omega_b$. Figure~\ref{fig3}(d), which magnifies the pink dashed region in Fig.~\ref{fig3}(c), shows that the optimal optomechanical entanglement occurs at $\Delta_m = 0.64\omega_b$ when $\kappa_m = 2\kappa_a$ (black curve). As the magnon decay rate increases ($\kappa_m = 5\kappa_a$ and $10\kappa_a$), the optimal point gradually shifts toward $\Delta_m \approx \omega_b$, indicating that stronger magnon damping alters the resonance condition of the hybrid modes. Finally, Fig.~\ref{fig3}(e) displays the magnon photon phonon tripartite entanglement $\mathcal{R}_\tau^{abm}$, which becomes pronounced around $\Delta_m \approx -\omega_b$. Increasing the magnon photon coupling significantly enhances this tripartite entanglement, implying that part of the optomechanical entanglement is converted into genuine three mode correlations among the Kittel mode, the cavity mode, and the mechanical mode.} 

\begin{figure}
	\includegraphics[scale=0.32]{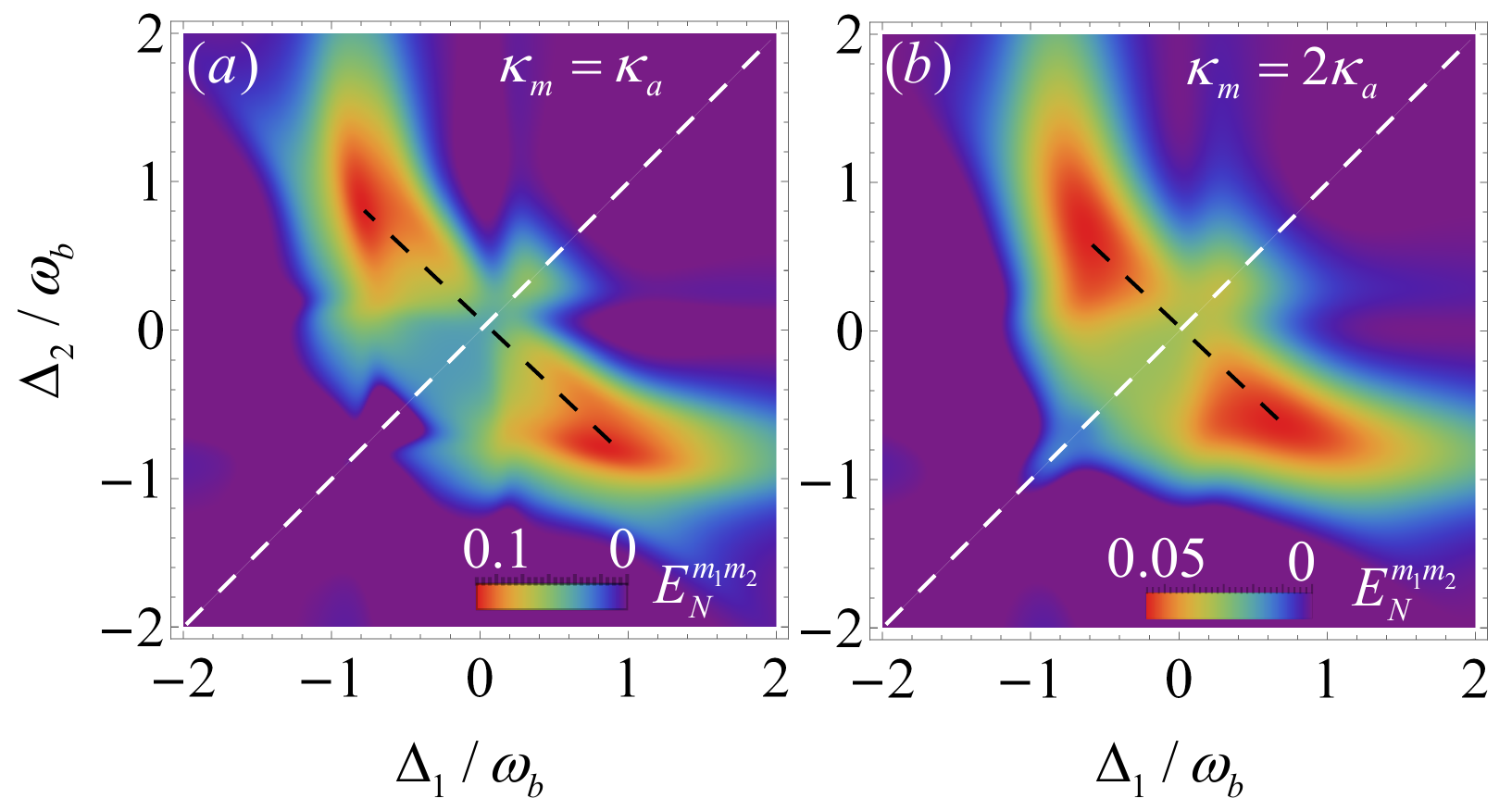} 
	\caption{Density plot of the magnon-magnon entanglement ($E_N^{m_1m_2}$) vs the normalized magnon frequency detunings $\Delta_1/\omega_b$ and $\Delta_2/\omega_b$ with (a) $\kappa_m=\kappa_a$ and (b) $\kappa_m=2\kappa_m$. Other parameters are the same as those in Fig.~\ref{fig2}(a)  except for $g_m=G=4\kappa_a$ and $\tilde{\Delta}_a=0.9\omega_b$.}\label{fig4}
\end{figure}

\section{Magnon-magnon entanglement with the optomechanical interface}\label{s4}

Based on the aforementioned study of the optomechanical entanglement, it is evident that the optomechanical interface can serve as an effective entanglement generator. Therefore, we proceed to demonstrate the generation of magnon-magnon entanglement within the framework of two spheres included cavity-magnon optomechanics (see Fig.~\ref{fig1}). 

\subsection{Magnon detuning effect}

Figure~\ref{fig4} plots the magnon-magnon entanglement ($E_N^{m_1m_2}$) vs the normalized frequency detunings of two Kittle modes of the spheres with $\kappa_m=\kappa_a$ and $\kappa_m=2\kappa_a$, where $g_m=G$ is fixed. The bat-shaped pattern in Fig.~\ref{fig4}(a) demonstrates that the magnon-magnon entanglement can be symmetrically adjusted by varying the detunings of the Kittle modes. By varying one detuning while keeping the other unchanged, we observe that the magnon-magnon entanglement initially increases to its optimal value, and then decreases to zero. The optimal magnon-magnon entanglement can be predicted around $\Delta_1=-\Delta_2\approx\pm\omega_b$, which are symmetrically distributed on both sides of the white dashed line $\Delta_1=\Delta_2$. The similar results are also evident from the bird-shaped pattern in Fig.~\ref{fig4}(b). By comparing Figs.~\ref{fig4}(a) and \ref{fig4}(b), we show that the larger magnon decay rate, the weaker magnon-magnon entanglement. Moreover, with increasing the magnon decay rate, the symmetric points about the diagonal line, where the optimal magnon-magnon entanglement is predicted, will converge in the case that magnons are resonant with the driving field, i.e., $\Delta_1=\Delta_2=0$.
\begin{figure}
	\includegraphics[scale=0.32]{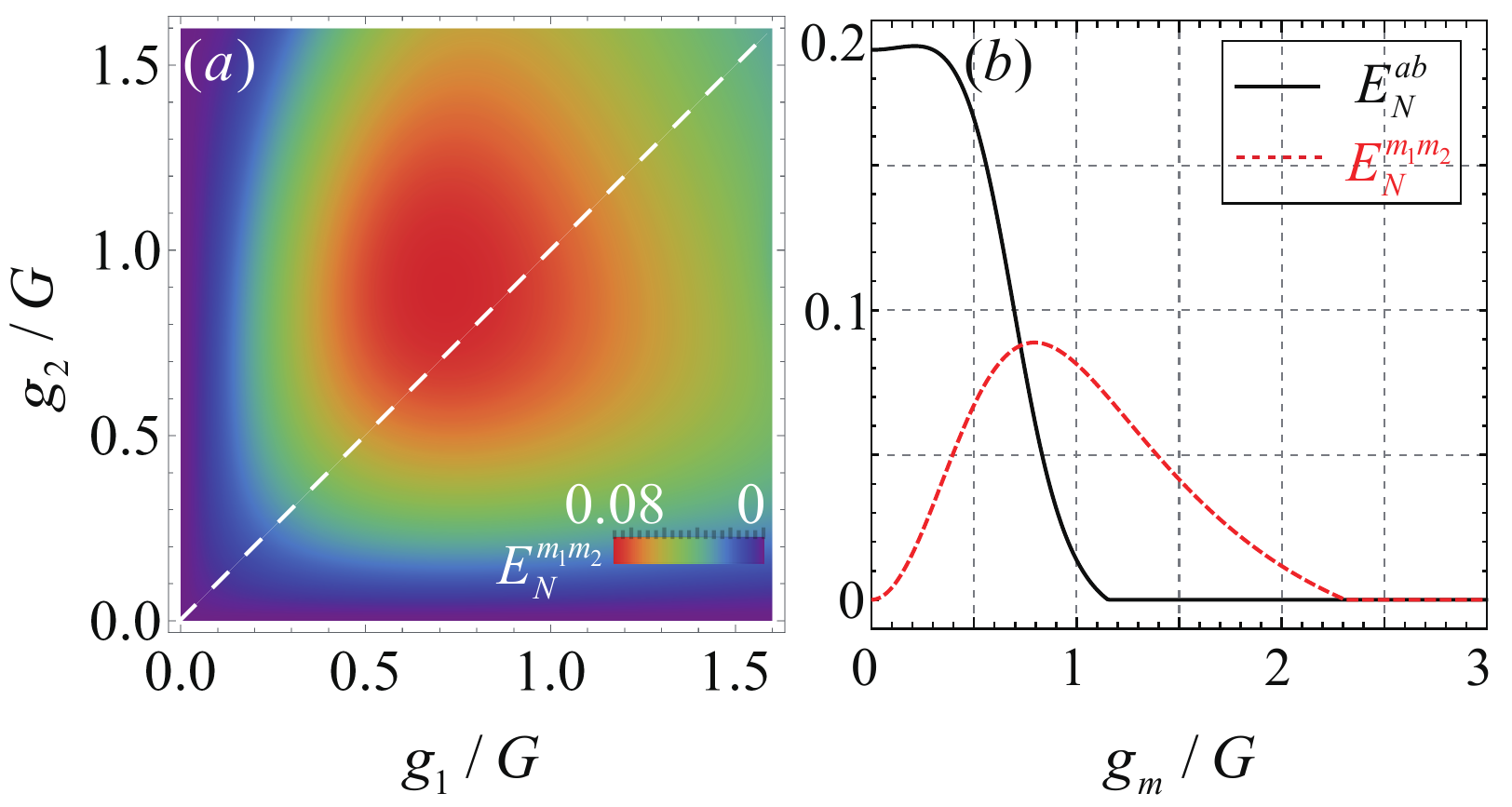} 
	\caption{(a) Density plot of the magnon-magnon entanglement ($E_N^{m_1m_2}$) vs the normalized magnon-photon coupling strengths $g_1/G$ and $g_2/G$. (b) The optomechanical entanglement ($E_N^{ab}$) and the magnon-magnon entanglement ($E_N^{m_1m_2}$) vs the normalized magnon-photon coupling strength $g_m/G$. Other parameters are the same as those in Fig.~\ref{fig2}(a) except for $\Delta_1=-\omega_b$, $\Delta_2=\omega_b$, $\kappa_m=\kappa_a$, $G=4\kappa_a$ and $\tilde{\Delta}_a=0.9\omega_b$.}\label{fig5g}
\end{figure}

\subsection{Magnon-photon coupling effect}

In this work, we assume that the magnons in the Kittle modes of two spheres have the identical coupling strength with the cavity mode. It is necessary to investigate the effect of two different coupling strengths $g_1$ and $g_2$ on the magnon-magnon entanglement generation. The numerical result is plotted in Fig.~\ref{fig5g}(a), where $\Delta_1=-\omega_b$ and $\Delta_2=\omega_b$ are fixed. Apparently, the optimal magnon-magnon entanglement can only be attained when two magnon-photon coupling strengths are comparable (see the white dashed line). With increasing the magnon-photon coupling strength, the magnon-magnon entanglement increases first to the optimal value, and then decreases to zero. To further show where the  magnon-magnon entanglement comes, we plot the magnon-magnon entanglemnt ($E_N^{m_1m_2}$) and the optomechanical entanglement ($E_N^{ab}$) vs the normalized coupling strenth $g_m/G$ in Fig~\ref{fig5g}(b). We find that, with increasing the magnon-photon coupling to $g_m \sim G$, the optomechanical entanglement decreases to the optimal value of the magnon-magnon entanglement. By further increasing the magnon-photon coupling, both the magnon-magnon entanglement and the optomechanical entanglement reduce to zero. Obviously, the former decreases much slower than the latter. The behavior in Fig.~\ref{fig5g}(b) reveals that the magnon-magnon entanglement is partially transfered from the optomechanical entanglement.

\subsection{Noise effect}

\begin{figure}
	\includegraphics[scale=0.32]{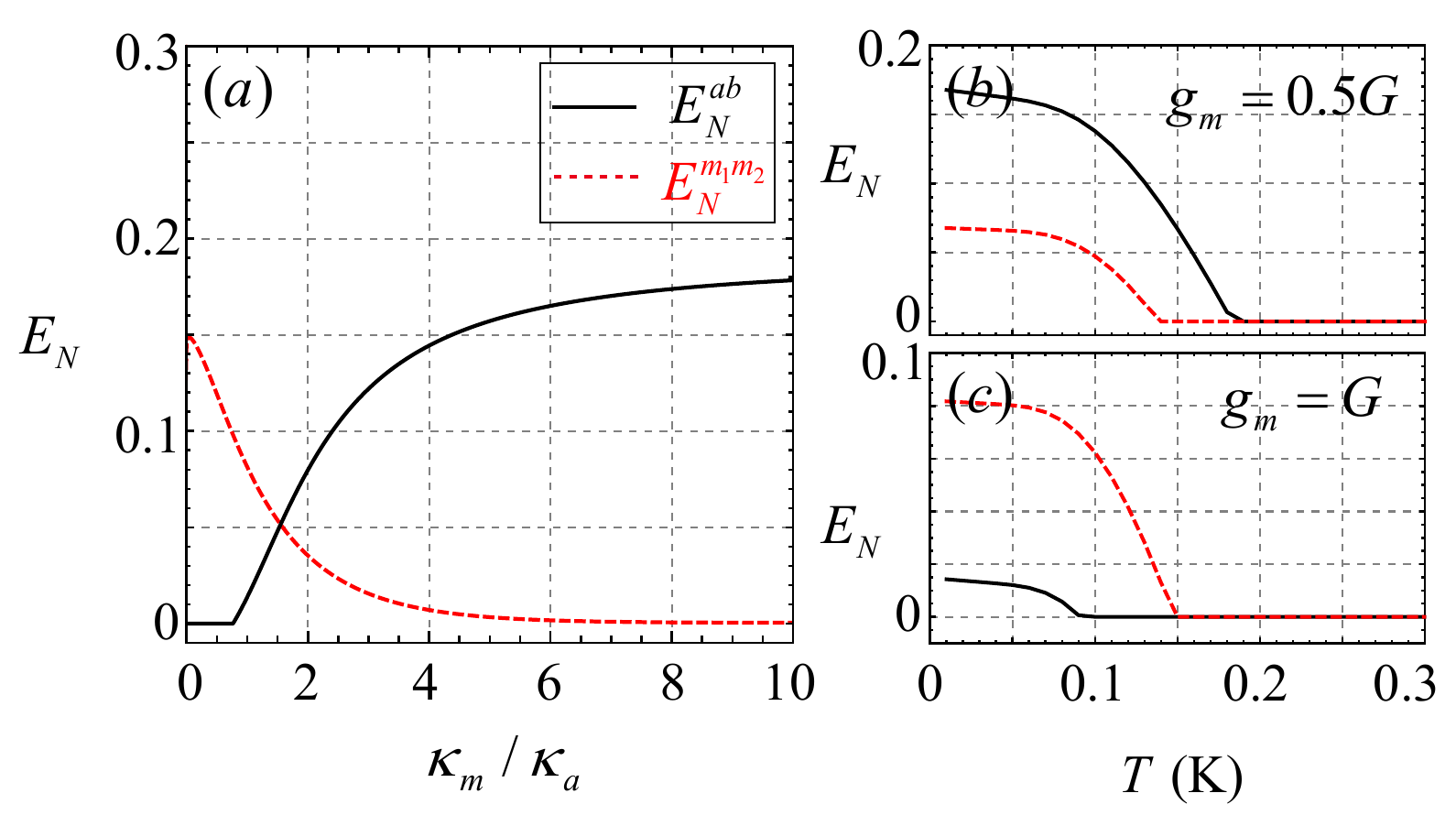} 
	\caption{The optomechanical entanglement ($E_N^{ab}$) and the magnon-magnon entanglement ($E_N^{m_1m_2}$) vs (a) the normalized decay rate $\kappa_m/\kappa_a$, and the bath temperature $T$ with (b) $g_m=0.5G$ and (c) $g_m=G$. Here $\Delta_1=-\omega_b$, $\Delta_2=\omega_b$, $G=4\kappa_a$,  and $\tilde{\Delta}_a=0.9\omega_b$ are fixed in (a-c). Other parameters are the same as those in Fig.~\ref{fig2}(a) except for (a) $g_m=G$,  (b,c) $\kappa_m=\kappa_a$.}\label{fig5}
\end{figure}

In Fig.~\ref{fig5}(a), the effects of the magnon decay rate on both the magnon-magnon and the optomechanical entanglement are investigated, where $g_m=G$ and $\Delta_1=-\Delta_2=-\omega_b$ are fixed. It is observed that larger decay rates of the magnons do not promote the generation of magnon-magnon entanglement. Instead, they facilitate the production of optomechanical entanglement, consistent with the findings from Figs.~\ref{fig3}(a) and \ref{fig3}(b). This not only reveals that the entanglement exchange between these two entanglement occurs, but also shows that the generated magnon-magnon entanglement can {be leveraged} to restore the optomechanical entanglement by increasing the decay rate of the Kittle mode.
Since the bath temperature is crucial for entanglement survival, we respectively plot $E_N^{m_1m_2}$ and $E_N^{ab}$ vs the bath temperature $T$ in Figs.~\ref{fig5}(b) and \ref{fig5}(c) with different magnon-photon coupling strengths, where $\kappa_m=\kappa_a$ is fixed. When the magnon-photon coupling strength is weaker than the optomechanical coupling strength [see $g_m=0.5G$ in Fig.~\ref{fig5}(b)], the optomechanical entanglement (the black curve) is stronger than the magnon-magnon entanglement (the red curve) within the survival temperature. But when the magnon-photon coupling strength is comparable with the optomechanical coupling strength [see $g_m=G$ Fig.~\ref{fig5}(c)], the situation is reversed, that is, the magnon-magnon entanglement becomes much stronger than the optomechanical entanglement. Moreover, the survival temperature of the optomechanical entanglement is significantly shifted with varying the magnon-photon coupling. This indicates that the survival temperature of the magnon-magnon entanglement is more robust against the change of the magnon-photon coupling than that of the optomechanical entanglement.

\section{Magnon-based tripartite entanglement}\label{s5}

Apart from the magnon-magnon entanglement, the optomechanical interface can also be used to produce tripatite entanglement, including the magnon-magnon-photon entanglement and the magnon-magnon-phonon entanglement, in hybrid cavity-magnon optomechanics.
\begin{figure}
	\includegraphics[scale=0.34]{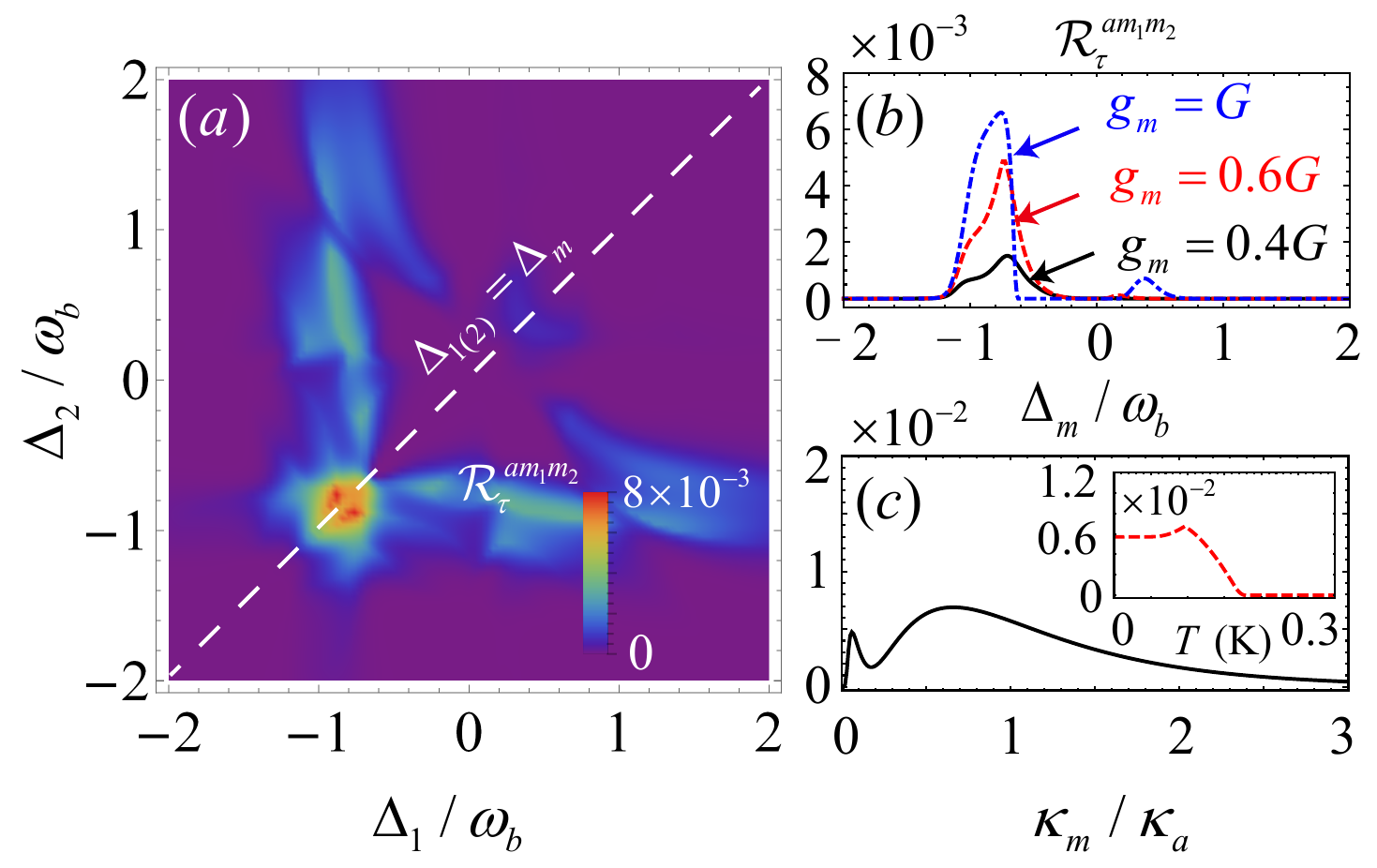} 
	\caption{(a) Density plot of the magnon-magnon-photon entanglement ($\mathcal{R}_\tau^{am_1m_2}$) vs the normalized frequency detunings $\Delta_1/\omega_b$ and $\Delta_2/\omega_b$, where $g_m=G$, $\kappa_m=\kappa_a$, and $T=20$ mK are taken. (b) The magnon-magnon-photon entanglement ($\mathcal{R}_\tau^{am_1m_2}$) vs the normalized frequency detuning $\Delta_m/\omega_b$ with different magnon-photon coupling strengths $g_m/G=0.4,0.6,1$, where $\kappa_m=\kappa_a$ and $T=20$ mK. (c) The magnon-magnon-photon entanglement ($\mathcal{R}_\tau^{am_1m_2}$) vs the normalized decay rate $\kappa_m/\kappa_a$, where $g_m=G$ and $\Delta_1=\Delta_2=-\omega_b$. The plot of the magnon-magnon-photon entanglement vs the bath temperature is shown in the inset. Other parameters are the same as those in Fig.~\ref{fig2}(a) except $G=4\kappa_a$  and $\tilde{\Delta}_a=0.9\omega_b$.}\label{fig6}
\end{figure}

\subsection{Magnon-magnon-photon entanglement}

In Fig.~\ref{fig6}(a), we plot the magnon-magnon-photon entanglement ($\mathcal{R}_\tau^{am_1m_2}$) vs the normalized detunings $\Delta_1/\omega_b$ and $\Delta_2/\omega_b$, where $g_m=G=4\kappa_a$, $\kappa_m=\kappa_a$, and $T=20$ mK. The eagle-pattern clearly indicates that the optimal magnon-magnon-photon entanglement is expected when magnons in two Kittle modes of the spheres are resonant and works on the blue-sideband, i.e., $\Delta_1=\Delta_2\approx-\omega_b$. When detunings deviate from this optimal point along the white diagonal line, where the resonant condition is maintained, we observe a rapid reduction in magnon-magnon-photon entanglement to zero. 
However, when nonidentical detunings are considered, this tripartite entanglement decreases slowly as detunings deviate from the optimal values, akin to a scenario where detunings stray from one of the eagle's wings. This indicates that nonidentical detunings can be used to protect this tripartite entanglement. 
Figure~\ref{fig6}(b) demonstrates that the magnon-magnon-photon entanglement can be significantly enhanced by appropriately increasing the magnon-photon coupling strengths while varying the detuning $\Delta_m=\Delta_1=\Delta_2$. Additionally, we observe that adjusting the magnon decay rate can also generate magnon-magnon-photon entanglement, as depicted in Fig.~\ref{fig6}(c). The inset reveals the robustness of magnon-magnon-photon entanglement against the bath temperature when $T<100$ mK. However, at a higher temperature, the entanglement diminishes and disappears around $T\approx180$ mK.

\begin{figure}
	\includegraphics[scale=0.34]{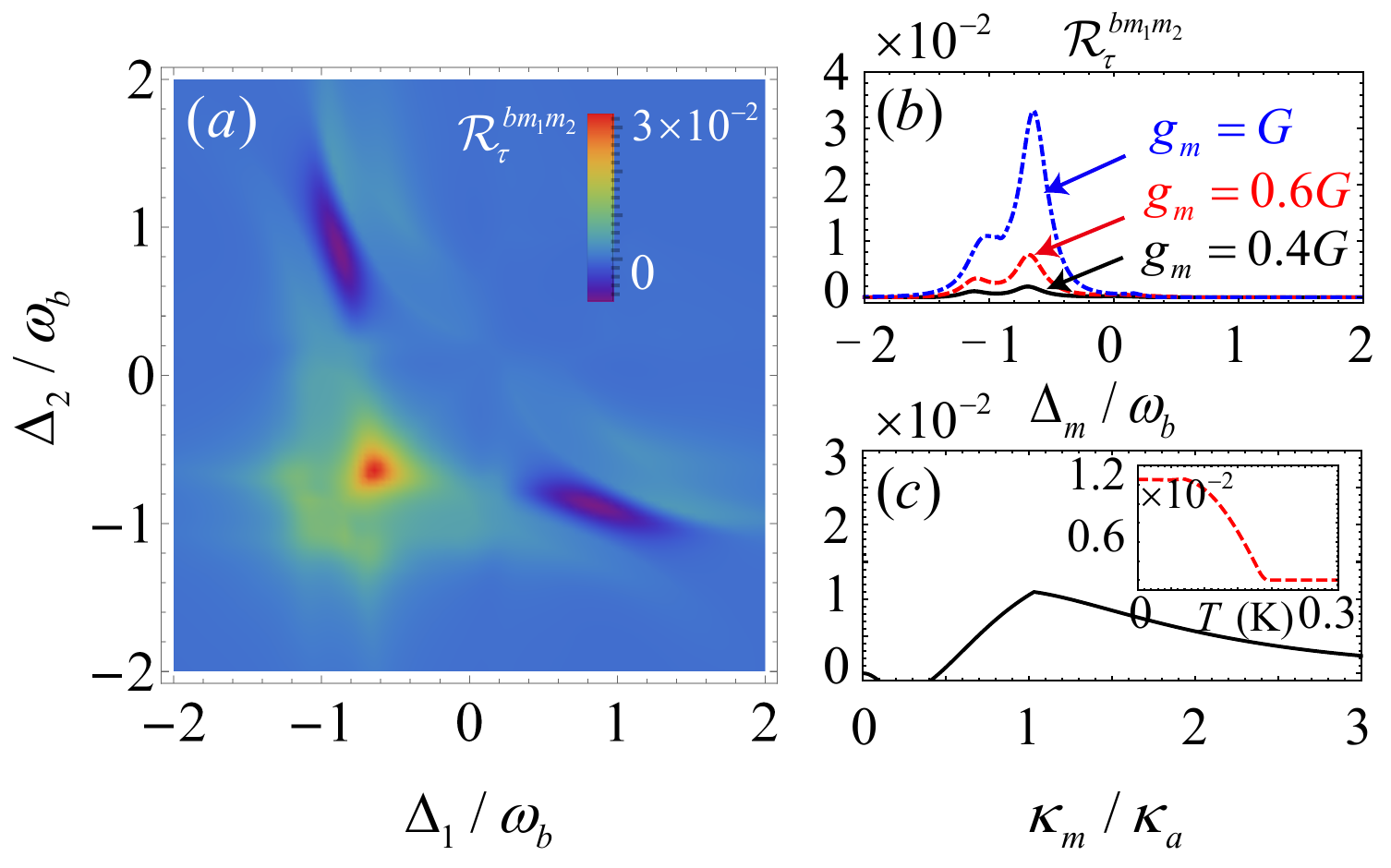} 
	\caption{(a) Density plot of the magnon-magnon-phonon entanglement ($\mathcal{R}_\tau^{bm_1m_2}$) vs the normalized frequency detunings $\Delta_1/\omega_b$ and $\Delta_2/\omega_b$, where $g_m=G$, $\kappa_m=\kappa_a$, and $T=20$ mK are taken. (b) The magnon-magnon-phonon entanglement ($\mathcal{R}_\tau^{bm_1m_2}$) vs the normalized frequency detuning $\Delta_m/\omega_b$ with different magnon-photon coupling strengths $g_m/G=0.4,0.6,1$, where $\kappa_m=\kappa_a$ and $T=20$ mK. (c) The magnon-magnon-phonon entanglement ($\mathcal{R}_\tau^{bm_1m_2}$) vs the normalized decay rate $\kappa_m/\kappa_a$, where $g_m=G$ and $\Delta_1=\Delta_2=-\omega_b/2$. The plot of the magnon-magnon-photon entanglement vs the bath temperature is shown in the inset. Other parameters are the same as those in Fig.~\ref{fig2}(a) except $G=4\kappa_a$ and $\tilde{\Delta}_a=0.9\omega_b$.}\label{fig7}
\end{figure}

\subsection{Magnon-magnon-phonon entanglement}

{Next, we investigate the magnon-magnon-phonon entanglement in the proposed hybrid cavity-magnon optomechanical system. It should be emphasized that there are no direct couplings between the two magnon modes or between the magnons and the phonon. Instead, the entanglement among them arises from indirect interactions mediated by the cavity field, which acts as an effective quantum bus~\cite{sarma2021continuous}. The optimal magnon-magnon-phonon entanglement is observed around $\Delta_1 = \Delta_2 \approx -0.5\omega_b$, corresponding to the tip of the fox-shaped pattern shown in Fig.~\ref{fig7}(a). Compared with Fig.~\ref{fig6}(a), the magnon-magnon-phonon entanglement is stronger than the magnon-magnon-photon entanglement. This can be attributed to two factors. First, the mechanical mode can be effectively cooled through the optomechanical beam-splitter interaction, which enhances its quantum correlations with the magnons \cite{aspelmeyer2012quantum, Bowen2015Book}. Second, the noise effect on the magnon-magnon-phonon entanglement is weaker because the magnons and phonons interact indirectly, which helps to suppress decoherence in this subsystem \cite{li2018magnon}.}

{Moreover, Fig.~\ref{fig7}(b) indicates that the tripartite magnon-magnon-phonon entanglement can be further enhanced by increasing the magnon-photon coupling strength, consistent with previous studies of hybrid quantum systems~\cite{qiu2022controlling,wang2015bipartite}. To explore the influence of the magnon decay rate, we plot the magnon-magnon-phonon entanglement ($\mathcal{R}_\tau^{bm_1m_2}$) as a function of the normalized decay rate $\kappa_m/\kappa_a$. The results show that a very small decay rate does not favor the formation of tripartite entanglement. As the decay rate increases, the entanglement first rises to a maximum and then gradually vanishes when losses dominate. Comparing Figs.~\ref{fig6}(c) and \ref{fig7}(c), both the magnon-magnon-phonon and magnon-magnon-photon entanglements exhibit similar resilience to the magnon decay rate. However, the magnon-magnon-phonon entanglement demonstrates stronger robustness against the bath temperature, as shown in the insets of Figs.~\ref{fig6}(c) and \ref{fig7}(c). Specifically, within the survival temperature range of approximately 180 mK, the magnitude of the magnon-magnon-phonon entanglement is nearly twice that of the magnon-magnon-photon entanglement. Note that the degree of tripartite entanglement achieved here remains weak. However, it could be further enhanced by introducing coherent feedback loop~\cite{amazioug2023enhancement}, parametric amplifier~\cite{hussain2022entanglement}, optical squeezing~\cite{zhang2022quantum}, combination of parametric and squeezing~\cite{wodedo2025optimizing}, mechanical nonlinearity~\cite{wan2024quantum}, and three-level atoms~\cite{kussia2024enhancement}. Incorporating these effects may provide an efficient route to strengthen multipartite quantum correlations, which will be an interesting topic for future investigation. To highlight the novelty of our scheme, a comparison with these existing schemes is provided in Table~\ref{table}.}

\begin{table}[htbp]
	\centering
	\caption{\textcolor{red}{Comparison of our work with others}}
	\label{table}
	\begin{adjustbox}{width=\columnwidth}
		\begin{tabular}{lcccc}
			\hline\hline
			\textbf{Reference} & \textbf{Method} & \textbf{Mechanism} & $\mathbf{E_N^{m_1 m_2}}$ & $\boldsymbol{\mathcal{R}_\tau^{m_1m_2a(b)}}$ \\
			\hline
			Ref.~\cite{amazioug2023enhancement} & Feedback loop & Magnomechanics & No & No \\
			Ref.~\cite{hussain2022entanglement} & Parametric amplifier & Magnomechanics & No & No \\
			Ref.~\cite{zhang2022quantum} & Squeezing & Magnomechanics & No & No \\
			Ref.~\cite{wodedo2025optimizing} & Parametric + squeezing & Optomechanics & No & No \\
			Ref.~\cite{wan2024quantum} & Mechanical nonlinearity & Optomechanics & No & No \\
			Ref.~\cite{kussia2024enhancement} & Three-level atoms & Optomechanics & No & No \\
			\textbf{Our work} & \textbf{Magnon} & \textbf{Optomechanics} & \textbf{Yes} & \textbf{Yes} \\
			\hline\hline
		\end{tabular}
	\end{adjustbox}
\end{table}

\section{conclusion}\label{s6}

{Before concluding, we establish the experimental feasibility of our hybrid system integrating a microwave optomechanical subsystem with a YIG sphere. This proposal leverages experimentally mature technologies: robust microwave optomechanical platforms in both 2D planar~\cite{blencowe2005nanoelectromechanical, greenberg2012nanomechanical, poot2012mechanical} and 3D bulk architectures~\cite{liu2025degeneracy, yuan2015large, gunupudi2019optomechanical}, alongside well-established $\mu$m-sized YIG spheres that exhibit strong coupling to microwave photons in magnonic systems~\cite{huebl2013high,tabuchi2014hybridizing,zhang2014strongly,ghirri2023ultrastrong,PhysRevAppliedHigh}. Crucially, the geometric compatibility challenge is resolved by scale disparity: 2D cavities (characteristic lengths $\sim$mm) accommodate $\mu$m-scale spheres due to orders-of-magnitude size difference~\cite{poot2012mechanical}, while 3D cavities with volumes $\gtrsim$100 mm³~\cite{yuan2015large} provide sufficient mode volume for microsphere integration. Consequently, the proposed setup is principly achievable with state-of-the-art techniques, though direct experimental demonstration of cavity-magnon optomechanics remains an open frontier.}

In summary, we propose to generate diverse bipartite and tripartite entanglement in hybrid cavity-magnon optomechanics. Unlike standard cavity optomechanics, our proposal not only enables magnon-mediated optomechanical entanglement but also allows for flexible magnon-magnon entanglement. It can even generate tripartite entanglement, including magnon-photon-phonon, magnon-magnon-photon, and magnon-magnon-phonon configurations. Moreover, optimal bipartite and tripartite entanglement can be achieved by tuning parameters. We further show that all entanglement can be enhanced via engineering the magnon-photon coupling, and are robust {below the survival threshold temperature}. In addition, we find that the optomechanical entanglement can be protected or restored by bad magnons with large decay rate, while other entanglement are significantly diminished. These findings suggest that our proposal offers a novel avenue to explore and control tunable macroscopic quantum effects in hybrid cavity-magnon optomechanics.

This work was supported by the Natural Science Foundation of Zhejiang Province (GrantNo. LY24A040004), Zhejiang Province Key R\&D Program of China (Grant No. 2025C01028),  and Shenzhen International Quantum Academy (Grant No. SIQA2024KFKT010).

\bibliography{ms}
\end{document}